\documentclass[11pt,a4paper]{article}
\usepackage[latin9]{inputenc}
\usepackage{amsmath}
\usepackage{amsthm}
\usepackage{amssymb}
\usepackage{graphicx}

\makeatletter

\usepackage{amsfonts}
\usepackage{amsthm}
\usepackage{epstopdf}

\newif\iffigs\figstrue

\makeatother

\begin{document}
\begin{titlepage} \vspace{0.3cm}
 \vspace{1cm}

\begin{center}
\textsc{\Large{}{}{}{}{}{}\ \\[0pt] \vspace{5mm}
 On the Existence of the Coleman Instantons 
}{\Large{}{}{}{}{}{} }\\[0pt] 
\par\end{center}


\begin{center}
\vspace{35pt}
 \textsc{V. F. Mukhanov$^{~a,b}$ and A. S. Sorin$^{~c,d,e}$}\\[15pt] 
\par\end{center}

\begin{center}
{$^{a}$ Ludwig Maxmillian University, \\[0pt] Theresienstr. 37,
80333 Munich, Germany\\[0pt] }e-mail: \textit{\small{}{}{}{}{}{}mukhanov@physik.lmu.de}{\small{}{}{}{}{}\vspace{10pt}
 }{\small\par}
\par\end{center}

 


\begin{center}
{$^{b}$ {\small{}{}{}{}{}{}Korea Institute for Advanced Study\\[0pt]
Seoul, 02455, Korea}}\vspace{10pt}
\par\end{center}


\begin{center}
{$^{c}$ {\small{}{}{}{}{}{}Bogoliubov Laboratory of Theoretical
Physics\\[0pt] Joint Institute for Nuclear Research \\[0pt] 141980
Dubna, Moscow Region, Russia \\[0pt] }}e-mail: \textit{\small{}{}{}{}{}{}sorin@theor.jinr.ru}{\small{}{}{}{}{}\vspace{10pt}
 }{\small\par}
\par\end{center}

 


\begin{center}
{$^{d}$ {\small{}{}{}{}{}{}National Research Nuclear University
MEPhI\\[0pt] (Moscow Engineering Physics Institute),\\[0pt] Kashirskoe
Shosse 31, 115409 Moscow, Russia}}\vspace{10pt}
\par\end{center}

\begin{center}
{$^{e}$ {\small{}{}{}{}{}{}Dubna State University, \\[0pt]
141980 Dubna (Moscow region), Russia}}\vspace{10pt}
\par\end{center}


\vspace{1cm}

\begin{center}
\textbf{{Abstract} } 
\par\end{center}

We identify infinite classes of potentials for which the Coleman instantons
do not exist. For these potentials, the decay of a false vacuum must
be described by the new instantons introduced in \cite{MRS1,MRS2}.

\newpage{}

Consider a scalar field with an unbounded potential $V\text{\ensuremath{\left(\varphi\right)}}$
such as shown in Fig.\ref{Figure1}. This potential has a local
minimum at $\varphi_{0}<0$ corresponding to a false vacuum, and
it is unbounded from below at positive $\varphi$. Obviously, the
false vacuum is unstable in this case and therefore must decay. Coleman
proposed \cite{Coleman} that the dominant contribution to the
decay rate comes from the $O\left(4\right)$-invariant Euclidean solution
of the scalar field equation 
\begin{equation}
\frac{\partial^{2}\varphi}{\partial t^{2}}-\Delta\varphi+\frac{dV}{d\varphi}=0\,,\label{eq:first}
\end{equation}
where the Minkowski time $t$ must be replaced by the Euclidean time
$\tau=\mathbf{i} \, t$. Then $O\text{\ensuremath{\left(4\right)}}$-invariant
solutions depend only on $\varrho=\sqrt{\tau^{2}+\mathbf{x^{2}}}$,
i.e. $\varphi\left(\tau,\mathbf{x}\right)=\varphi\left(\varrho\right)$
and equation (\ref{eq:first}) reduces to the ordinary second-order
differential equation 
\begin{equation}
\ddot{\varphi}(\varrho)+\frac{3}{\varrho}\,\dot{\varphi}(\varrho)-V'=0\,,\label{eq:2}
\end{equation}
where $V'\equiv dV/d\varphi$ and $\dot{\varphi}\equiv d\varphi/d\varrho$. Two boundary conditions must be added
to this equation. One of them is obvious, namely, 
\begin{equation}
\varphi\left(\varrho\rightarrow\infty\right)=\varphi_{0}\,.\label{eq:3}
\end{equation}
Coleman imposed the second condition 
\begin{equation}
\dot{\varphi}\left(\varrho=0\right)=0
\label{eq:4}
\end{equation}
to avoid the singularity as $\varrho\rightarrow0$, because otherwise
the action 
\begin{equation}
S_{I}\,=\,2\pi^{2}\,\int_{0}^{+\infty}d\varrho\,\varrho^{3}\,\left(\frac{1}{2}\,\dot{\varphi}^{2}\,+\,V(\varphi)\right)\label{Eeq1a}
\end{equation}
could diverge and thus the contribution of the singular instanton
to the decay rate, which is proportional to $\exp\left(-S_{I}\right)$,
would vanish.

\vspace{0.5cm}


\begin{figure}[hbt]
\begin{centering}
\includegraphics[height=60mm]{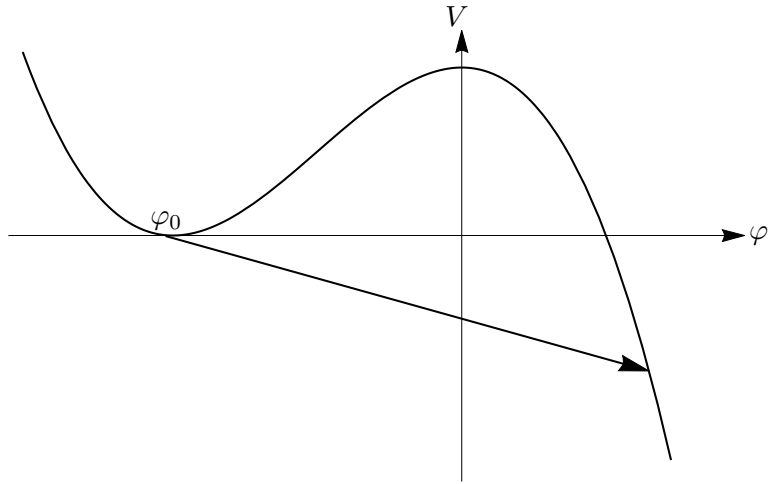} 
\par\end{centering}
\vspace*{-3.9cm}
 \hspace*{5.1cm} $\varphi_{0}$

\vspace*{-3.1cm}
 \hspace*{9.1cm}$V$

\vspace*{2.35cm}
 \hspace*{13.1cm}$\varphi$

\vspace*{3cm}
\caption{The unbounded potential for which the false vacuum at $\varphi_{0}$
must be unstable.}
\label{Figure1} 
\end{figure}


In \cite{CGM}, it was found under what conditions on the potential
$V$ the Coleman instantons exist with certainty. The purpose of this
note is somehow complimentary. We want to find a whole class of unbounded
potentials for which the Coleman instantons, which must simultaneously
satisfy two boundary conditions (\ref{eq:3}) and (\ref{eq:4}),
do not exist.

To do this, we first integrate equation (\ref{eq:2}) and find
the following set of nonlocal integrals of motion: 
\begin{align}
E\left(\alpha\right) & =\varrho^{\frac{4}{\alpha-2}}\left(\frac{1}{2}\varrho^{2}\,\dot{\varphi}^{2}+\frac{2}{\alpha-2}\,\varrho\,\varphi\,\dot{\varphi}-\varrho^{2}\,V-\frac{2\left(\alpha-4\right)}{\left(\alpha-2\right)^{2}}\,\varphi^{2}\right)\nonumber \\
 & +\frac{2}{\alpha-2}\int_{0}^{\varrho}d\overline{\varrho}\,\overline{\varrho}^{\frac{6-\alpha}{\alpha-2}}\left[\left(\alpha-4\right)\left(\overline{\varrho}\,\dot{\varphi}+\frac{2}{\alpha-2}\,\varphi\right)^{2}+\overline{\varrho}^{2}\left(\alpha\,V-\varphi\,V'\right)\right]\,\label{eq:5}
\end{align}
parameterized by the parameter $\alpha$.
To verify that this expression is indeed the integral of motion for
any value of $\alpha$, one can differentiate $E\left(\alpha\right)$
with respect to $\varrho$, 
\begin{equation}
\frac{dE}{d\varrho}=\varrho^{\frac{\alpha+2}{\alpha-2}}\left(\ddot{\varphi}(\varrho)+\frac{3}{\varrho}\,\dot{\varphi}(\varrho)-V'\right)\left(\varrho\,\dot{\varphi}+\frac{2}{\alpha-2}\varphi\right)\,,\label{eq:6}
\end{equation}
which obviously vanishes on-shell of equation (\ref{eq:2}).

First, we note that the expression on the right-hand side of (\ref{eq:5})
does not change if we add a constant to the potential $V$.
Therefore, without loss of generality we can normalize $V$ so
that $V(\varphi=0)=0$. We assume that the potential at
$\varphi<0$ has the deepest local minimum (false vacuum) at $\varphi_{0}<0,$
where $V\left(\varphi_{0}\right)<0$, and otherwise it is completely
arbitrary (see Fig.\ref{Figure2}). For positive $\varphi$, the
potential is unbounded from below, and we will determine for which
such unbounded potentials the Coleman instantons do not exist.

\vspace{0.5cm}


\begin{figure}[hbt]
\begin{centering}
\includegraphics[height=60mm]{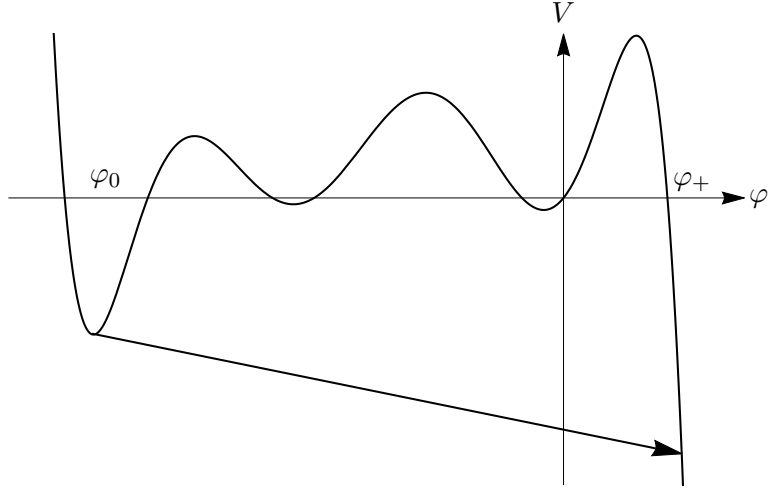} 
\par\end{centering}
\vspace*{-4.5cm}
 \hspace*{4.3cm} $\varphi_{0}$

\vspace*{-2.6cm}
 \hspace*{10.5cm}$V$

\vspace*{1.9cm}
 \hspace*{13.1cm}$\varphi$

\vspace*{-0.7cm}
 \hspace*{12.1cm}$\varphi_{+}$

\vspace*{4cm}
\caption{Since the integral (\ref{eq:5}) does not depend on the constant term
in the potential, for convenience we normalized the potential so that
$V(\varphi=0)=0$.}
\label{Figure2} 
\end{figure}


Let us assume that the field tunnels from the false vacuum at $\varphi_{0}<0$
to $\varphi_{+}>0$. For the Coleman instanton, this tunneling can
be thought of as the ``motion of the particle'' with friction in the
inverted potential $-V$ (see equation (\ref{eq:2})). This ``motion''
starts at $\varphi_{+}$ at $\varrho=0$ with zero velocity and ``the
particle'' moves towards $\varphi_{0}$, which should be reached
as $\varrho\rightarrow\infty$, if the Coleman instanton exists. If
the ``initial condition'' (\ref{eq:4}) is satisfied, then the integral
of motion $E\left(\alpha\right)$ (\ref{eq:5}) at $\varrho=0$
for $\alpha>2$ must obviously be equal to zero, i.e. 
\begin{equation}
E(\alpha)=0\,.\label{Ealpha}
\end{equation}
Suppose that the field moving in the direction of negative $\varphi$
vanishes at finite $\varrho_{0}$, i.e. $\varphi\left(\varrho_{0}\right)=0$.
Then expression (\ref{eq:5}) for this $\varrho_{0}$ and $\alpha>2$
takes the following form: 
\begin{equation}
E\left(\alpha\right)=\frac{1}{2}\,\varrho_{0}^{\frac{2\alpha}{\alpha-2}}\dot{\varphi}(\varrho_{0})^{2}+\frac{2\left(\alpha-4\right)}{\alpha-2}\int_{0}^{\varrho_{0}}d\overline{\varrho}\,\overline{\varrho}^{\frac{6-\alpha}{\alpha-2}}\left(\overline{\varrho}\,\dot{\varphi}+\frac{2}{\alpha-2}\,\varphi\right)^{2}+\frac{2}{\alpha-2}\int_{0}^{\varrho_{0}}d\overline{\varrho}\,\overline{\varrho}^{\frac{\alpha+2}{\alpha-2}}\,\varphi^{\alpha+1}\,v'_{\alpha}\,,\label{eq:6a}
\end{equation}
where the rescaled potential $v_{\alpha}\left(\varphi\right)$ is
introduced instead of $V\left(\varphi\right)$: 
\begin{equation}
V\left(\varphi\right)\equiv -\varphi^{\alpha}\,v_{\alpha}\left(\varphi\right)\,.\label{eq:6b}
\end{equation}
For $\alpha\geq4$ the sum of two first terms on the right-hand side of
equation (\ref{eq:6a}) is positive and the third term is non-negative if 
\begin{equation}
v'_{\alpha}\equiv\frac{d\,v_{\alpha}\left(\varphi\right)}{d\varphi}\geq 0\,.\label{eq:7}
\end{equation}
In this case $E(\alpha)>0$ regardless of the ``initial'' positive
value of $\varphi(\varrho=0)$, which contradicts equation (\ref{Ealpha})
required by the Coleman boundary condition (\ref{eq:4}). This implies
that the field $\varphi(\varrho)$ cannot cross $\varphi=0$ and proceed
to move towards $\varphi_{0}$, therefore, it remains positive for all
finite values of $\varrho$ \footnote{In fact, even if $\dot{\varphi}$($\varrho=0)$ is singular but does
not diverge as fast as $\varrho^{-\frac{\alpha}{\alpha-2}}$, the
solution always remains in the range of positive $\varphi$.}. Thus, for any unbounded potential, which at positive $\varphi$
can equivalently be represented as 
\begin{equation}
V(\varphi)=-\varphi^{\alpha}\int^{\varphi}d\bar{\varphi}\,v'_{\alpha}\left(\bar{\varphi}\right)+\varLambda\,\varphi^{\alpha}\label{eq:8a}
\end{equation}
with $\alpha\geq4$ and $v'_{\alpha}\geq0$, the Coleman instanton
describing the decay of the deepest false vacuum at $\varphi_{0}<0$ with $V(\varphi_0)<0$
does not exist regardless of the form of the potential at negative
$\varphi$, where $\varLambda$ is the integration constant.

Let us consider a few concrete examples.

\textbf{a) }Taking first $v'_{\alpha}=0$ in (\ref{eq:8a}), we find
that for any unbounded potential decreasing at positive $\varphi$
as $V=-\lambda\,\varphi^{\alpha}$ with arbitrary, not necessarily
integer $\alpha\geq4$, the Coleman instanton does not exist. As can
be seen from (\ref{eq:6a}), the value of $E(\alpha)$ at $\varphi=0$
must be strictly positive if $\alpha>4$, which contradicts $E(\alpha)=0$
imposed by (\ref{eq:4}), and hence the field $\varphi$ never reaches
$\varphi=0$ at finite $\varrho$. In the special case of the $-\lambda\,\varphi^{4}$-
potential, we have $\dot{\varphi}=0$ at $\varphi=0$. The corresponding
solution describing the tunneling from the top of the $-\lambda\,\varphi^{4}$-potential
to a positive $\varphi$ is called the Fubini--Lipatov instanton
\cite{F,L}. This solution cannot be continued to $\varphi<0$ regardless
of what the potential is at negative $\varphi$. Therefore, strictly
speaking, the Fubini--Lipatov instanton cannot describe the instability
in the Standard Model contrary to numerous claims in the literature.

Taking $\alpha=4$ and $v'_{4}=\lambda/\varphi>0$ in (\ref{eq:8a}),
one can conclude that for the one loop modified unbounded potential
\begin{equation}
V=-\frac{\lambda}{4}\,\varphi^{4}\ln\left(\frac{\varphi}{\mu}\right)\label{eq:9a}
\end{equation}
the Coleman instanton, which is supposed to describe the decay of
the false vacuum at $\varphi_{0}<0$, does not exist either. In the
case of the approximated Standard Model 2-loop effective potential
\cite{SM1}, the situation is more complicated and will be discussed
in \cite{MS-SM}.

\textbf{b) }For 
\begin{equation}
v'_{\alpha}=\lambda\,\varphi^{\beta-\alpha-1}\,,\label{10a}
\end{equation}
where $\lambda>0$, we obtain from (\ref{eq:8a}) the following potential
\begin{equation}
V=-\frac{\lambda}{\beta-\alpha}\,\varphi^{\beta}+\varLambda\,\varphi^{\alpha}\,.\label{eq:11-1}
\end{equation}
Here, the first term is negative and dominates when $\beta>\alpha\geq 4$.
Therefore, the Coleman instanton does not exist, regardless of the
sign of the constant $\Lambda$, which can be either negative or positive.
For $\beta<\alpha$ the potential is unbounded only if $\varLambda<0$.
In this case, the first term in (\ref{eq:11-1}) is positive. For
example, for $V=-\lambda\,\varphi^{4}+m^{2}\varphi^{2}$ at $\varphi\geq0$
the instanton is absent only if $m^{2}\geq0$, contrary to naive
expectations. Let us note that both $\alpha\geq4$ and $\beta$ in
(\ref{eq:11-1}) are not necessarily integer.

\textbf{c) }Finally, we construct an infinite class of polynomials
\begin{equation}
V(\varphi)=\sum_{i=1}^{N}a_{i}\,\varphi^{i},\label{eq:12a}
\end{equation}
for which inequality (\ref{eq:7}) is satisfied and thus the Coleman
instantons do not exist. First, we set $\alpha=4$ and consider 
\begin{equation}
v'_{4}=\varphi^{-4}\,a\,\left(\varphi^{2}+b\,\varphi+c\right).\label{eq:13a}
\end{equation}
The function $v'_{4}$ must be positive at all $\varphi>0$ and therefore
the quadratic polynomial in the parenthesis can be represented in
the form 
\[
a\left(\varphi-\lambda_{1}\right)\left(\varphi-\lambda_{2}\right)
\]
with positive $a$, while the two roots $\lambda_{1}$ and $\lambda_{2}$
must be either negative or complex conjugate. Let us assume
that they are complex conjugate, i.e. $\lambda_{1,2}=-\beta\pm{\mathrm{i}}\,\gamma$.
In this case, 
\begin{equation}
v'_{4}=a\,\varphi^{-4}\left(\left(\varphi+\beta\right)^{2}+\gamma^{2}\right)\label{eq:14a}
\end{equation}
is positive for any $\beta$ and $\gamma$. By substituting (\ref{eq:14a})
into (\ref{eq:8a}), we find that the unbounded potentials 
\begin{equation}
V\left(\varphi\right)=\varLambda\,\varphi^{4}+a\left(\varphi^{3}+\beta\,\varphi^{2}+\frac{\beta^{2}+\gamma^{2}}{3}\,\varphi\right)\label{eq:15a}
\end{equation}
with $\varLambda<0$ and $a\geq0$ do not have the Coleman instantons.
Let us note that here the mass term can be either negative, positive,
or zero. Similarly, we can consider the case when both roots $\lambda_{1}$
and $\lambda_{2}$ are negative and we present the result for completeness:
\begin{eqnarray}
V\left(\varphi\right)=\varLambda\,\varphi^{4}+a\left(\varphi^{3}-\frac{\lambda_{1}+\lambda_{2}}{2}\,\varphi^{2}+\frac{\lambda_{1}\,\lambda_{2}}{3}\,\varphi\right)\,.\label{eq:7c}
\end{eqnarray}

In the same way one can derive the polynomial potential (\ref{eq:12a})
with an arbitrary integer power $N\equiv\alpha\geq4$ that
satisfies inequality (\ref{eq:7}). Taking 
\[
v'_{\alpha}=\varphi^{-\alpha}\,a\,\left(\varphi^{\alpha-2}+b\,\varphi^{\alpha-3}+...+c\right)\,,
\]
one has to determine when $v'_{\alpha}$ is positive for $\varphi>0.$
This is only when $a>0$ and the polynomial in the brackets
has $2m$ $(m=1\,,2\,,...\,,[\frac{N}{2}])$ complex conjugate roots
and the remaining $\alpha-2-2m$ real roots are negative. Thus, the
explicit \textit{unique general solution} that satisfies inequality
(\ref{eq:7}) with the full moduli space of its parameters corresponds
to 
\begin{eqnarray}
v'_{\alpha}=a\,\varphi^{-\alpha}\prod_{i=1}^{\alpha-2\,(m+1)}\left(\varphi-\lambda_{i}\right)\prod_{j=1}^{m}\left(\left(\varphi+\beta_{j}\right)^{2}+\gamma_{j}^{2}\right),\label{eq:7a}
\end{eqnarray}
where all $\gamma_{j}\,,\lambda_{i}\leq0$ and $\beta_{j}$ are arbitrary.
Substituting this expression into (\ref{eq:8a}) and taking simple
indefinite integrals, we find explicitly polynomial potentials
for which instantons with the Coleman boundary conditions do not exist.

Similarly, one can show that inequality (\ref{eq:7}) is satisfied
for the potentials (\ref{eq:8a}) with arbitrary values of $\alpha\geq4$ and
\begin{eqnarray}
 v'_{\alpha}(\varphi)=a\prod_{i=1}^{N-2\,(m+1)}\left(\varphi-\lambda_{i}\right)\prod_{j=1}^{m}\left(\left(\varphi+\beta_{j}\right)^{2}+\gamma_{j}^{2}\right)\,,\label{eq:7ab}
\end{eqnarray}
where the integer parameter $N\neq\alpha$, so that for these potentials the Coleman instantons
do not exist as well.

Thus, we have constructed infinite classes of unbounded potentials
for which the false vacuum must obviously be unstable, but still its
decay cannot be described by the Coleman instantons with the boundary
conditions (\ref{eq:3}--\ref{eq:4}). We argued in \cite{MRS1,MRS2}
that the inevitable quantum fluctuations make the boundary condition
(\ref{eq:4}) formulated in the deep ultraviolet regime completely
unreliable and therefore it must be abandoned. These quantum fluctuations
introduce the ultraviolet cutoff determined by the parameters of the
classical solution, beyond which this solution cannot be trusted.
The cutoff in turn regularizes the solutions that were previously
singular, leading to the appearance of a broad class of new instantons
\cite{MRS1,MRS2}. These new instantons must necessarily be used to
describe the instability of the false vacuum for the potentials considered
in this work.

\bigskip{}

\textbf{Acknowledgments}


The work of V. M. was supported by the Germany Excellence Strategy---EXC-2111---Grant
No. 39081486.

The work of A. S. was supported in part by RFBR grant No. 20-02-00411.


\end{titlepage} 
\end{document}